\documentstyle[epsf,11pt]{article}
\newcommand{\bea}{\begin{eqnarray}}
\newcommand{\eea}{\end{eqnarray}}

\begin{document}
\begin{titlepage}

\vspace{0.2cm}

\title{{\normalsize \hfill HIP-2000-32/TH}\\[20pt]
R-parity violation in $(t+\bar t )\tilde g$ production at LHC
and Tevatron \footnote{The authors thank the Academy of Finland
(project number 163394) for financial support.}}
\author{M. Chaichian$^{a,b}$, K. Huitu$^a$ and Z.-H. Yu$^{a,b}$\\
$^a$Helsinki Institute of Physics\\
$^b$Department of Physics, University of Helsinki\\
P.O.Box 9, FIN-00014 University of Helsinki, Finland}
\date{}
\maketitle

\vspace*{2truecm}

\begin{center}\begin{minipage}{5in}

\begin{center} ABSTRACT\end{center}
\baselineskip 0.2in

{We study the production of $(t+\bar{t}) \tilde{g}$ at the hadron colliders
in an R-parity ($R_{p}$) violating supersymmetric model. 
This process provides us with information not only about $R_{p}$ violation,
but may also help us in detecting the supersymmetry itself.
It is possible to detect an $R_{p}$ violating
signal (with single gluino production) at the future hadron colliders, 
such as Fermilab Tevatron Run II or CERN Large Hadron Collider (LHC), 
if the parameters in the supersymmetric $\rlap/\! R_{p}$ interactions
are not too small, e.g. for $m_{\tilde{g}}=1$ TeV, 
$\lambda^{''}=0.1$, still
hundreds of events are produced at LHC with luminosity $30\ fb^{-1}$.
Even if we could not detect a signal of $\ \rlap/\!R_{p}$ in the
experiment, we
get stringent constraints on the heavy flavour $\rlap/\!R_{p}$ 
couplings. 
In addition to the minimal supersymmetric standard model we have  
also considered some models with a heavy gluino as the lightest
supersymmetric particle.
} \\

\vskip 5mm

{~~~~PACS number(s): 13.65.+i, 13.88.+e, 14.65.-q, 14.80.Dq, 14.80.Gt}
\end{minipage}
\end{center}
\end{titlepage}

\baselineskip=0.3in

\eject
\rm
\baselineskip=0.36in
\baselineskip=0.18in

\begin{flushleft} {\bf I. Introduction} \end{flushleft}
\par
It is well known that the minimal supersymmetric standard model (MSSM)
\cite{s1} in its most general form contains lepton number ($L$) 
and baryon number ($B$) violating couplings.
The resulting catastrophic proton decay \cite{s2} can be avoided by imposing
R-parity symmetry $R_p$,
\begin{equation}
R_{p}=(-1)^{3B+L+2S},
\end{equation}
where $S$ is the spin of the particle.
In the models with R-parity conservation, superparticles can only be pair 
produced and the lightest supersymmetric particle (LSP) will be stable. 
However, $R_{p}$ conservation is not necessary for forbidding proton decays,
instead of that, we just need either B-conservation or L-conservation
\cite{s3}. 
Models with $R_{p}$ violation ($\rlap /\! R_{p}$)
can provide many interesting 
phenomena, such as neutrino masses and mixing. Partly because of that,
$\rlap /\!R_{p}$ has attracted much attention \cite{s4,s5,s6}.
Many constraints from low-energy phenomenology \cite{s6}
are collected in Ref. \cite{s7}. 

The supersymmetry (SUSY) must be broken, since it has not been
observed so far. 
Two kinds of breaking mechanisms of supersymmetry have been
extensively studied phenomenologically, namely the minimal
supergravity (mSUGRA) \cite{s8} and minimal gauge-mediated SUSY
breaking (GMSB) \cite{s9}. 
These predict a different pattern of masses especially for the
partners of gauge bosons, gauginos.
Thus, finding the signals of gauginos is an important way to probe SUSY.
In mSUGRA and GMSB, 
it is assumed that masses of gauginos will be unified at Grand Unified
Theory (GUT) scale. 
From the evolution of parameters, gluinos should be
the heaviest gauginos at low scale, since the ratios of gaugino
masses to coupling constants do not change with scale
in one-loop approximation \cite{s10}. 
Thus gluinos can decay to other
gauginos with jets in the $R_{p}$-conserving model \cite{a1}.

However, heavy gluino as LSP (or NLSP) may still 
exist in some GMSB models, as S. Raby suggested in \cite{s11}.
He introduced the Higgs-Messenger mixing in GMSB model.
This would lead triplet-doublet messengers split from 
triplet-doublet Higgs splitting. Since gluino obtains mass from
SUSY-breaking induced by triplet messengers, heavier triplet messengers
could suppress the mass of
gluino so that gluino could be the LSP (or NLSP). 
The experimental limits on the masses of gluinos have been discussed 
in $R_{p}$-conserving model \cite{s12}. 
In $R_{p}$-violating model, gluinos can decay through $R_{p}$-violating
channels, which obviously changes both the detection strategy 
and present mass limits \cite{s13}.

In the high energy hadron colliders especially squarks and gluinos
will be produced plentifully, if they are not too heavy. 
It is hoped for that information on SUSY is found in Fermilab Tevatron Run II,
or in the future CERN Large Hadron Collider, where squarks and
gluinos with masses below 1.5 TeV should be detected \cite{s14}. 
In mSUGRA and GMSB models, masses of squarks and gluinos are usually
of the same order.
However, in some special mechanism, such as O-II model \cite{s15},
it has been suggested that squarks can be much heavier than gauginos. 
In this model the gluinos will be produced at much lower energies than
squarks, so that it can be the first detected SUSY particle
at hadron colliders.
Since gluinos would be almost degenerate
with the lightest neutralino and chargino, $R_{p}$ violating
decay of gluinos will be significant \cite{s7}.

The single production of gluinos, neutralinos and charginos
has already been considered in the general case \cite{a2}, and
the single production of squarks, which is also significant 
to detect SUSY and $R_{p}$-violation,
has been considered recently \cite{t1}. 
In this work we will consider $\rlap /\!R_{p}$ at hadron colliders
in the process 
\bea
PP(P\bar{P}) \rightarrow t+\tilde{g}\,
 (\bar{t}+\tilde{g}).
\eea
This process occurs via 
B-violating terms in the $\rlap /\!R_{p}$ model.
In terms inducing heavy flavours, the $\rlap /\!R_{p}$ 
couplings can be very large from the present upper limits \cite{s7}.
For example, $\lambda^{''}_{2ij}$ and $\lambda^{''}_{3ij}$, getting their
strongest constraints from the ratio of widths of $Z$ to leptons and hadrons,
can be of order one (${\cal{O}}$(1)) for the  sfermion mass
${\cal{O}}$(100 GeV).

The pair production and decay of gluinos at hadron colliders
have already been researched in Ref. \cite{s16}. 
It has been shown that detecting gluinos is very difficult.
It may become easier with an accompanying top
quark in the process which we consider. 

On the other hand, the single top quark production is an interesting
topic itself \cite{s17}.
The single top quark production in the $\rlap /\!R_{p}$ model
has been considered in Ref. \cite{t2}. There, the possible
cross section will depend on the $\rlap /\!R_{p}$ parameters
as $|\lambda^{''}|^{4}$, while the process we will consider
depends on $|\lambda^{''}|^{2}$. 
Although the mass of the gluino is not known presently,
it is still possible to get
stronger constraints on $R_{p}$-violating parameters from the 
$(t+\bar{t}) \tilde{g}$ production.
\par
In the following, 
we will give the analytical
calculations of $PP(P\bar{P}) \rightarrow t \tilde{g} (\bar{t} \tilde{g})$
in section II.
In section III gluino and top decays are considered and in section IV
the numerical results are presented. 
The conclusions are given in section V and some details of the expressions
are listed in the appendix.


\begin{flushleft} {\bf II. Production of gluinos in 
$PP\rightarrow t\tilde g$ } \end{flushleft}

The superpotential for $\rlap/\! R_{p}$ violating, but gauge and
supersymmetry preserving  interactions is written as \cite{s6}
$$
\begin{array} {lll}
    W_{\rlap/\! R_{p}} & =
\lambda_{[ij]k} L_{i}.L_{j}\bar{E}_{k}+\lambda^{'}_{ijk}
L_{i}.Q_{j}\bar{D}_{k}+ \lambda^{''}_{i[jk]}
\bar{U}_{i}\bar{D}_{j}\bar{D}_{k}+\epsilon _{i} L_{i} H_{u}
\end{array}
\eqno {(2.1)}
$$
where $L_i$, $Q_i$ and $H_u$ are SU(2) doublets containing lepton, quark
and Higgs superfields respectively, $\bar{E}_j$ ($\bar{D}_j$, $\bar{U}_j$)
are the singlet lepton superfields (down-quark and up-quark).
The square brackets around the generation indices $i,j$ denote
antisymmetry of the bracketted indices.

We ignored the last term in Eq. (2.1) because
its effects are assumed small in our process \cite{s7}.
Thus, we have 9 $\lambda$-type, 27 $\lambda^{'}$-type and 9
$\lambda^{''}$-type independent parameters left.
The constraints on the couplings \cite{s7},
$$
|(\lambda~{\rm or}~\lambda^{'}) \lambda^{''}|<10^{-10}
\left(\frac{\tilde{m}}{100
\,{\rm GeV}}\right)^{2}.
\eqno {(2.2)}
$$
is usually taken to indicate that only L- or B-number violating
couplings exist.

In our work we will only consider the baryon number violating
couplings, i.e., the third term in Eq. (2.1).
In the following calculations we assume the parameters 
$\lambda^{''}$ to be real. 

We define the Mandelstam variables as usual
$$
    s  = (p_{1}+p_{2})^2=(p_{3}+p_{4})^2,
\eqno {(2.3.a)}
$$
$$
    t  = (p_{1}-p_{3})^2=(p_{4}-p_{2})^2,
\eqno {(2.3.b)}
$$
$$
    u  = (p_{1}-p_{4})^2=(p_{3}-p_{2})^2.
\eqno {(2.3.c)}
$$

\begin{figure}[t]
\leavevmode
\begin{center}
\mbox{\epsfxsize=8.truecm\epsfysize=5.truecm\epsffile{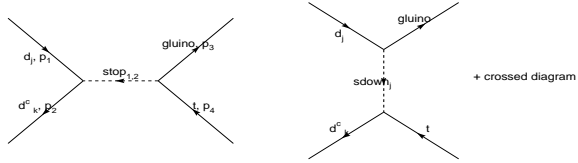}}
\end{center}
\caption{\label{feyn}Feynman diagrams of 
$q_{j}q_{k}\rightarrow \bar{t} \tilde{g}$.}
\end{figure}
The amplitude (Feynmann diagrams in Fig.1) of 
$q_{j}q_{k}^{\prime} \rightarrow \bar{t} \tilde{g}$ 
is given by:
$$
\begin{array}{lll}
M&=& M_{s}+M_{t}+M_{u},
\end{array}
\eqno{(2.4)}
$$
with
$$
\begin{array}{lll}
M_{s}=& &\overline{u^{a}} (p_3) (-i \sqrt{2} g_{s} 
T^{a}_{\alpha \alpha^{\prime}})
(\cos \theta P_{L}+\sin \theta P_{R}) v_{\alpha^{\prime}} (p_4)
\frac{i}{(s-m_{\tilde{t}_{1}}^{2}+im_{\tilde{t}_1}\Gamma_{\tilde{t}_1})} \\
&\times& \overline{v^{c}_{\gamma}} (p_2) 
(2i \epsilon^{\alpha\gamma\beta} \lambda^{''} _{3kj})
\sin \theta P_{R} u_{\beta} (p_1)\\
&+&\overline{u^{a}} (p_3) (-i \sqrt{2} g_{s} T^{a}_{\alpha \alpha^{\prime}})
(\sin \theta P_{L}-\cos \theta P_{R}) v_{\alpha^{\prime}} (p_4)
\frac{i}{(s-m_{\tilde{t}_{2}}^{2}+im_{\tilde{t}_2}\Gamma_{\tilde{t}_2})} \\
&\times& \overline{v^{c}_{\gamma}} (p_2) (-2i \epsilon ^{\alpha\gamma\beta} 
\lambda^{''} _{3kj})
\cos \theta P_{R} u_{\beta} (p_1) ,\\
\end{array}
\eqno{(2.5.a)}
$$
$$
\begin{array}{lll}
M_{t}&= &\overline{u^{a}} (p_3) (i \sqrt{2} g_s T^{a}_{\beta \rho} P_{R})
u_{\rho} (p_1)
\frac{i}{(t-m_{\tilde{d}}^{2})} 
\overline{u^{c}_{\alpha}} (p_4) 
(-2i \epsilon^{\alpha\beta\gamma} \lambda^{''} _{3jk} P_{R})
u_{\gamma} (p_2),\\
\end{array}
\eqno{(2.5.b)}
$$
$$
\begin{array}{lll}
M_{u}&= &\overline{u^{a}} (p_3) (i \sqrt{2} g_s T^{a}_{\beta \gamma} P_{R})
u_{\gamma} (p_2)
\frac{i}{(u-m_{\tilde{d}}^{2})} \overline{u^{c}_{\alpha}} (p_4) 
(-2i \epsilon^{\alpha\beta\rho} \lambda^{''} _{3kj} P_{R})
u_{\rho} (p_1),\\
\end{array}
\eqno{(2.5.c)}
$$
where $P_{L,R}$ are left- and right-helicity projections respectively,
$\theta$ is the mixing angle of stop quarks (see Appendix for details) and
$\Gamma_{\tilde{t}_{1,2}}$ are decay widths of stop quarks
$\tilde{t}_{1,2}$.
The families of down-type quarks are marked by 
$j,k=1,2,3$ and the upper index $c$ means 
charge conjugate. In the calculations, we have neglected all mixing
angles of scalar quarks except stop quarks.  
The amplitude depends on the $R_{p}$-violating parameters
$\lambda^{''}_{3jk}$ ($j,k=1,2,3$) in the process.

The total cross section for the
process $q_{j}q_{k}^{'} \rightarrow \bar{t} \tilde{g}$ is:
$$
\begin{array}{l}
\hat{\sigma}(\hat{s}) = \frac{1}{16 N_{c}^2 \pi \hat{s}^2 }
             \int_{\hat{t}^{-}}^{\hat{t}^{+}} d\hat{t}\, 
              {\overline{\sum}_{spins}^{}}\, |M|^{2},
\end{array}
\eqno{(2.6)}
$$
where $\hat{t}^{\pm}=\frac{1}{2}\left[
(m_t^2+m_{\tilde{g}}^2-\hat{s})\pm \sqrt{\hat{s} ^2+m_t^4+m_{\tilde{g}}^4-2\hat{s} m_t^2-2\hat{s}
m_{\tilde{g}}^2-2 m_t^{2} m_{\tilde{g}}^{2}} \right]$
and $M$ is the amplitude.
Here we have neglected the masses of
incoming quarks. $N_{c}=3$ is the color factor and the bar over summation means
averaging over the initial spins.

In a similar way, the cross section for 
$\bar{q}_{j}\bar{q}_{k}^{'} \rightarrow t \tilde{g}$ can be calculated.
The possible effects of 
$q_{j}q_{k}^{'} \rightarrow \bar{t} \tilde{g}$ and
$\bar{q}_{j}\bar{q}_{k}^{'} \rightarrow t \tilde{g}$ should be observed in $P\bar{P}$ or $PP$ colliders. 
The cross section for the process $P(P_{1})P(P_{2}) \rightarrow
(t + \bar {t}) \tilde{g} X$
 can be obtained by convoluting
the subprocess with quark distribution functions \cite{s20},
$$
\begin{array} {lll}
\sigma (s) &= \int dx_{1} dx_{2} f_{i}(x_{1},Q) f_{j}(x_{2},Q)
              \hat{\sigma}(\hat{s},\alpha_{s}(\mu))
\end{array}
\eqno {(2.7)}
$$
with $p_{1}=x_{1}P_{1}$, $\tau=x_{1}x_{2}=\hat{s}/s$.
$f_{i,j}(x_{n},Q)\, (n=1,2)$ are the corresponding quark distribution 
functions of protons. We take $Q=\mu=300$ GeV. 
Similarly we can find numerically the cross
section of 
$P(P_{1})\bar{P}(P_{2}) \rightarrow (t + \bar {t}) \tilde{g}
X^{\prime}$. 


\begin{flushleft} {\bf III. Top and gluino decays} \end{flushleft}

In the process of Eq. (2), we have in the
final state two heavy particles which will possibly decay inside the
detector.
Here we will shortly review the relevant decay modes of the
top quark and gluino \cite{toplhc,cdem,bel,hm,s13}.

The top  decays in the MSSM have been considered by several 
authors \cite{toplhc,cdem,bel,hm,hollik}.
The main decay mode is $t\rightarrow bW$, but 
$t\rightarrow bH^{+}$ can compete with it if mass of the charged
Higgs is lighter than $m_t-m_b$.
Top quark decay to R-odd particles
will also be important if those superparticles are light
enough \cite{hollik}. However, in our case, with heavy squarks, the
decays to real superparticles are impossible, except for 
the light LSP gluino
with light squark \cite{s12}. We plot the ratio $\Gamma_{\tilde{t}_1
\tilde{g}}:\Gamma_{bW}$ as a function of the stop quark 
$\tilde{t}_1$ mass, $m_{\tilde t_1}$, in Fig.2 for the gluino mass 
$m_{\tilde{g}}=30$ GeV.
In order to guarantee
the purported standard top quark events at the Tevatron, 
$BR(t\rightarrow bW)$ should be larger than $40-50\%$ as lower
bound \cite{topb}.
So with the assumption of light LSP gluino (about $25 \ - \ 35$ GeV),
lower limit on the mass of stop quark can be obtained.  
For the top quark decay through $R_p$-violating interactions \cite{bel},
the 
branching ratio of those decay modes will be very small compared
with $t\rightarrow bW$ in our case.
The only decay channel which we will consider in detecting top quark
is the $t\rightarrow bW\rightarrow bl\nu_l$, where $l=e,\mu$.
We will confine us to these decays (with $BR(W\rightarrow l\nu_l)\sim
22\%$, its branching ratio in top quark decay should be at least
$8.8-11\%$) 
since it is assumed that they have less background and are thus easier to
detect than the hadronic decay modes.

\begin{figure}[t]
\leavevmode
\begin{center}
\mbox{\epsfxsize=7.truecm\epsfysize=5.truecm\epsffile{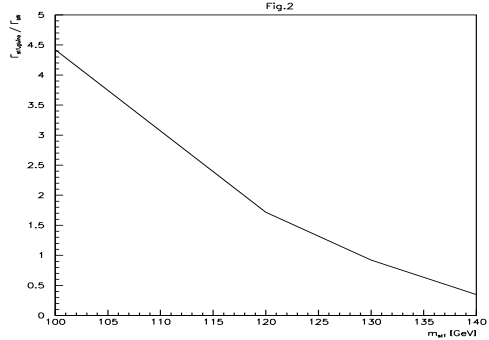}}
\end{center}
\caption{\label{feyn}ratio of $\Gamma_{\tilde{t}_1
\tilde{g}}:\Gamma_{bW}$ as function of mass of stop quark $\tilde t_1$,
with $m_{\tilde g}= 30\ {\rm GeV}$.}
\end{figure}

We consider in this work a heavy gluino, which may be the LSP.
The decay modes of a heavy gluino have been looked at
e.g. in \cite{s13,s14} in $R_p$ violating case.
The decay channel $\tilde g \rightarrow q \tilde q$ will dominate if
kinematically allowed.
A lighter than squarks gluino will decay to 
\bea
\tilde g\rightarrow q\bar q\tilde\chi^0_i,\, q\bar q'\tilde\chi^\pm_j,\,
 g\tilde\chi^0_i,
\eea
where $\tilde\chi^0_i$ and $\tilde\chi^\pm_i$ are neutralinos and
charginos, respectively.
The R-parity breaking decay modes become important for large
$\lambda^{''}$: 
\bea
\tilde g\rightarrow  q_i q_j q_k.
\eea
In Fig.3, we draw the $\tilde g$ decay with parameters in mSUGRA
model.
We consider the decay of gluinos in Fig.3 (a) and (b) with
$R_{p}$ conservation and violation, respectively.
In our calculations, only the two lightest neutralinos and
the lightest chargino are considered, since the other neutralinos
and the heavier chargino are too heavy for gluino to decay into.
It is shown that the gluino branching ratio to two heavy quarks (top quark
or
bottom quark) and neutralinos or charginos (or jets from
$R_p$-violating interactions) can be very large
and increases with mass of gluino.
This is reasonable
because stop quarks and sbottom quarks can be much lighter
than the other scalar quarks due to the large Yukawa couplings.
Especially, in $R_{p}$-violating terms, the process through virtual
stop quark $\tilde{t}_{1}$ will
dominate the decay width because in the mSUGRA model it is the lightest
squark and its mixing angle is near $\pi/2$.
Thus, there are two $b$ (or $t$) quarks as a signal of gluino
in our process (branching ratio can be read from
Fig.3, as about $ 60-70\% $).
Combined with other top quark,
the three heavy quarks, leads to the final state
$3 b + n (l+\nu)$ ($n=1,2,3$).
This final state can be detected in the future CERN LHC
and distinguished from background (with an assumed 
b-tagging efficiency $\epsilon \sim 50 \%$ in LHC \cite{was}).
For the much heavier gluinos which can decay directly to top quark
and stop quark, decay width is shown in Fig.3 (c).
\begin{figure}[t]
\leavevmode
\begin{center}
\mbox{\epsfxsize=5truecm\epsfysize=5truecm\epsffile{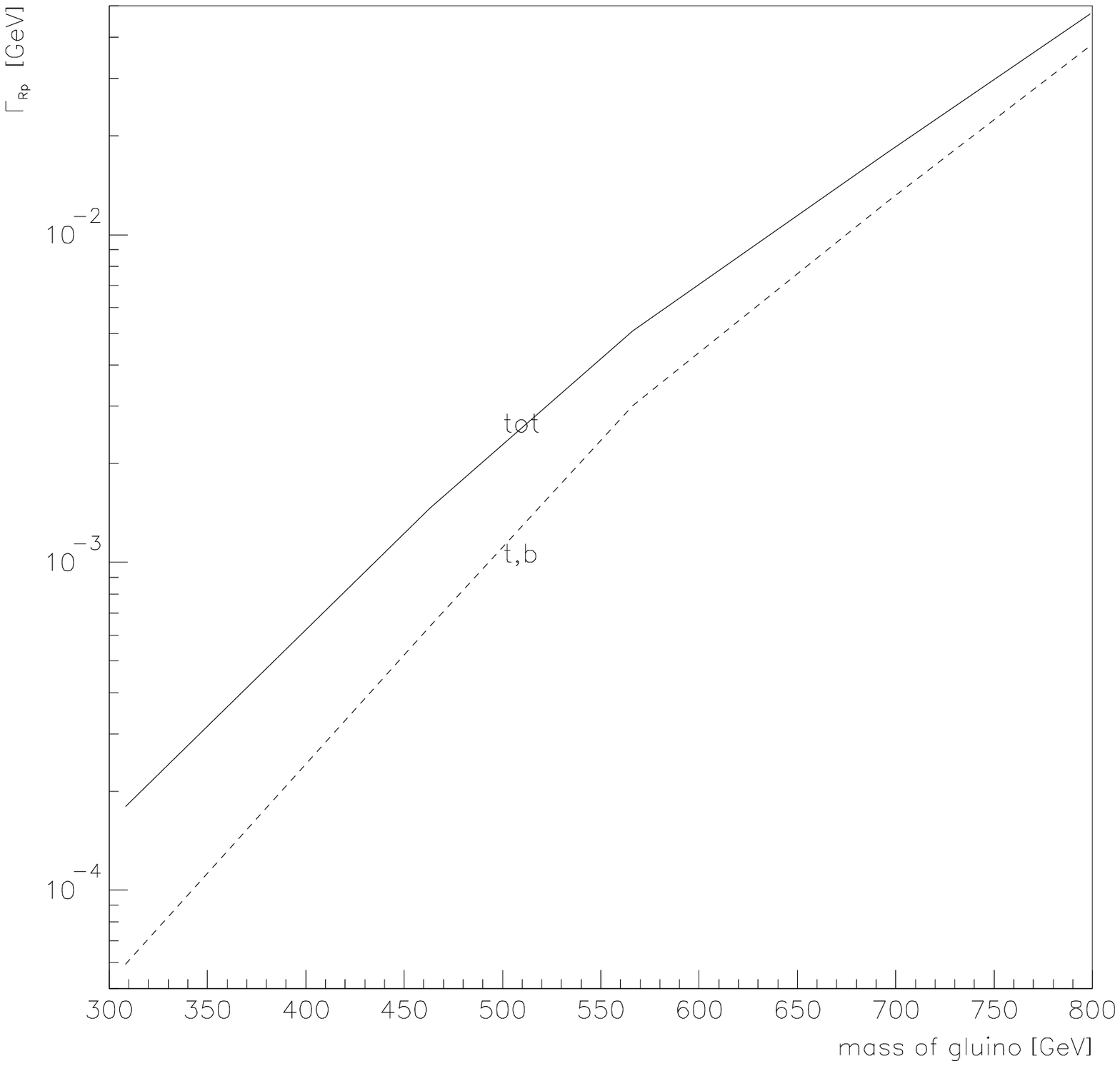}}
\mbox{\epsfxsize=5truecm\epsfysize=5truecm\epsffile{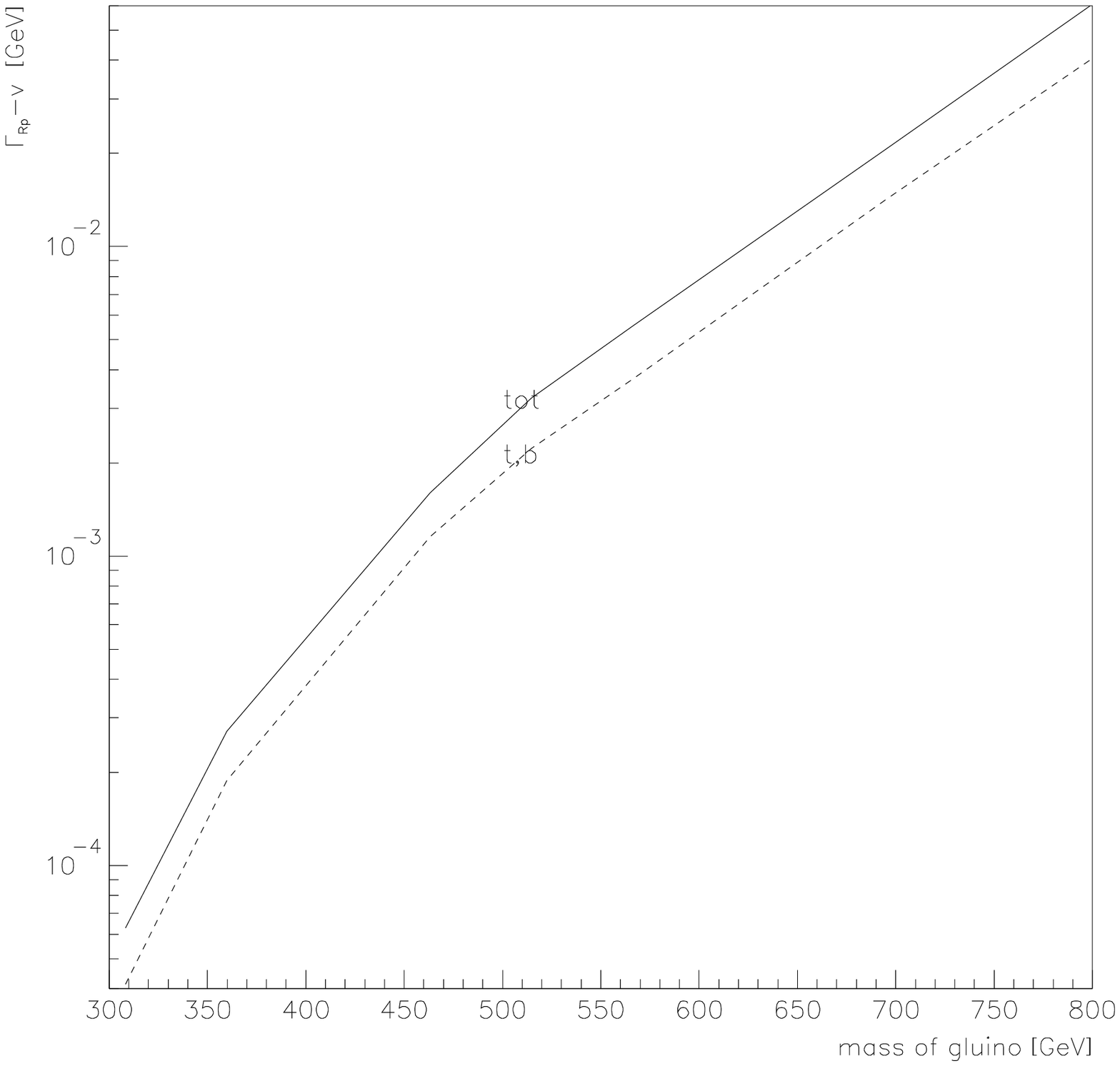}}
\mbox{\epsfxsize=5truecm\epsfysize=5truecm\epsffile{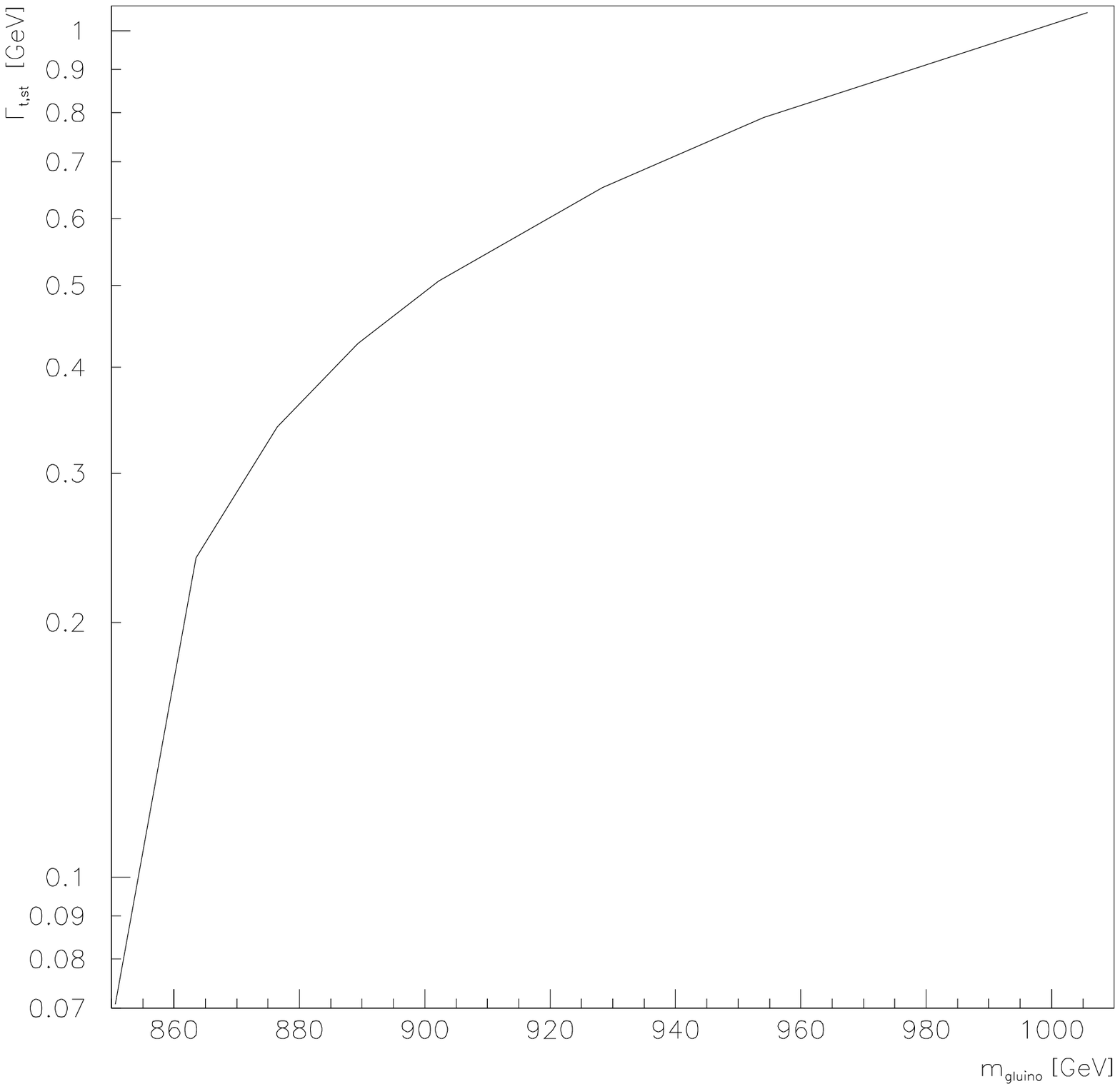}}
\end{center}
\caption{\label{gldecay}a) Decay width of gluino
               as a function of mass of gluino,
               $m_{\tilde{g}}$, for $R_{p}$ conservation.
                        Solid line corresponds to the total decay width
and
                        dashed line to partial width containing heavy
                        quarks (top quark and bottom quark).
b) Decay width of gluino
               as a function of
               $m_{\tilde{g}}$ for $R_{p}$ violation.
                        Solid line corresponds to the total decay
width and
                        dashed line to the partial width containing heavy
                        quarks (top and bottom).
c) Decay width of gluino to $t\tilde{t}_1$
               as a function of $m_{\tilde{g}}$.}
\end{figure}

When  gluino is the LSP, the only available decay modes are the
R-parity violating ones.  
If the R-parity violating couplings are not exceedingly small,
the gluino will decay in the detector through these channels.
However, if the couplings are very small, the gluinos will form
so-called R-hadrons before decaying (see e.g. \cite{s11,s12}).

\begin{flushleft} {\bf IV. Numerical results} \end{flushleft}


In the mSUGRA model we are interested in the region of the parameter
space where the gluino is lighter than the squarks, with the
possible exception of the lighter stop quark.
This choice allows gluino to be produced via the third generation
$R_p$-breaking coupling with relatively large cross section and on 
the other hand it is not complicated by
gluino decay to other squarks than possibly the lightest stop.

As a representative example of this part of the parameter space,
we take
$m_0=1000\,{\rm GeV}, A_0=-1000\,{\rm GeV}, \tan \beta=10$ and 
$sign(\mu)=+1$.
Varying $m_{1/2}$ suitably gives us the relevant gluino masses.
The resulting masses for supersymmetric particles with the varied
$m_{1/2}$ are listed in Table 1\footnote{We thank A. Wodecki for 
providing us his program, which calculates the sparticle masses in MSSM.
The program checks also some phenomenological constraints.}.

\begin{center} {\bf Table.1} MSSM parameters with $m_{0}=1$ TeV,
                                $A_0=-1$ TeV, $\tan \beta =10$ and
                                $sign(\mu)=+$, units of values are GeV in
the table.
\par
\begin{tabular}{|c|c|c|c|c|c|c|c|} \hline
$m_{\frac{1}{2}}$ &$ m_{\tilde{g}}$ & $m_{\tilde{t}_1}$ &$ m_{\tilde{t}_2}$ &
$m_{\tilde{b}_1}$ &$ m_{\tilde{q}}$ &$ m_{\tilde{\chi}^0_1}$ &
$m_{\tilde{\chi}^{\pm}_1}\sim m_{\tilde{\chi}_2^0}$ \\ \hline
120 & 308 & 457 &813 & 790 & 1020-1040 & 49 & 95 \\
\hline
140 & 360 & 468 &828&803&1028-1055& 58&111 \\ \hline
180 & 463 & 499 &862&836&1060-1090&75&144 \\ \hline
200&515&517&882&855&1075-1110&83&161 \\ \hline
220&566&537&904&876&1100-1130&92&177 \\ \hline
250&643&571&939&910&1120-1160&104&202 \\ \hline
280&721&608&977&947&1160-1200&117&227 \\ \hline
300&773&634&1004&974&1190-1230&125&243 \\ \hline
330&850&675&1047&1016&1220-1270&138&268 \\ \hline
340&876&689&1061&1031&1240-1280&142&277 \\ \hline
350&902&703&1076&1045&1250-1300&146&285 \\ \hline
370&954&733&1106&1076&1280-1330&155&302 \\ \hline
390&1006&763&1138&1107&1310-1360&163&318 \\ \hline
\end{tabular}
\end{center}


\begin{figure}[t]
\leavevmode
\begin{center}
\mbox{\epsfxsize=6.truecm\epsfysize=6.truecm\epsffile{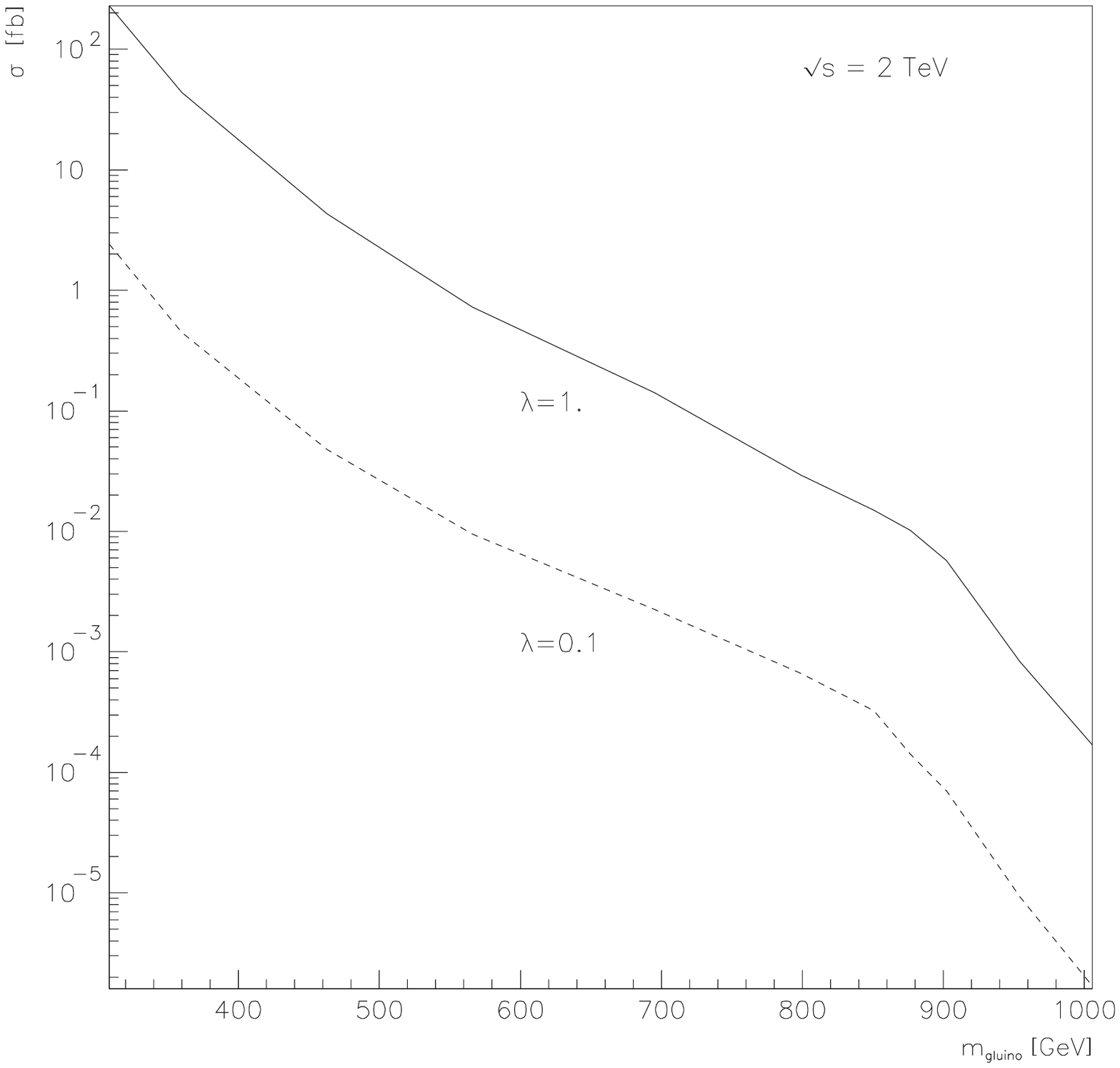}}
\mbox{\epsfxsize=6.truecm\epsfysize=6.truecm\epsffile{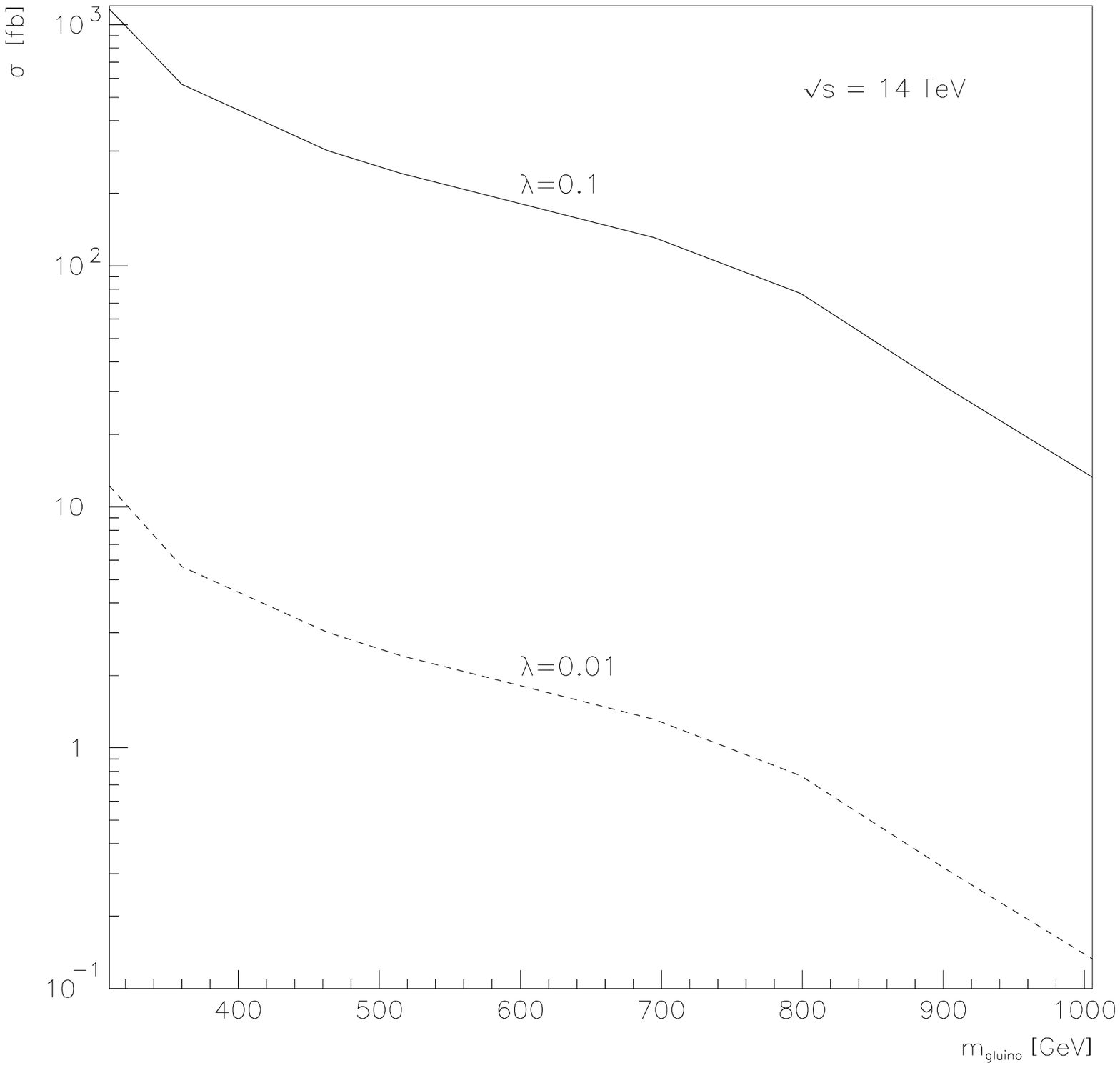}}
\end{center}
\caption{
{\label{cs1}} a) Cross section of 
$P\bar{P} \rightarrow (t+\bar{t}) \tilde{g} X$ as a function of 
$m_{\tilde{g}}$ with   $E_{CMS}=2$ TeV.
Solid line corresponds to $\lambda^{''}_{3ij}=1.0$ and dashed line to 
$\lambda^{''}_{3ij}=0.1$.  b) Cross section of $PP \rightarrow 
(t+\bar{t}) \tilde{g} X$ as a function of $m_{\tilde{g}}$ with 
$E_{CMS}=14\, {\rm TeV}, \lambda^{''}_{3ij}=0.1$.}
\end{figure}

In Fig.4 (a), we show the cross section of $P\bar{P}\rightarrow
(t+\bar{t}) \tilde{g} X$ as a function of mass of gluino
($m_{\tilde{g}}$) at Tevatron Run II energy, i.e.
with center-of-mass energy of the collision $\sqrt{s}=2$ TeV.
There, the solid line corresponds to 
$\lambda^{''}_{312}=\lambda^{''}_{313}=\lambda^{''}_{323}=1$
and dashed line to $\lambda^{''}_{312}=\lambda^{''}_{313}
=\lambda^{''}_{323}=0.1$.
At LHC with $\sqrt{s}=14$ TeV, the cross section for $PP\rightarrow
(t+\bar{t}) \tilde{g} X$ as a function of $m_{\tilde{g}}$
is shown in Fig.4 (b) with
$\lambda^{''}_{312}=\lambda^{''}_{313}=\lambda^{''}_{323}=0.1$
for solid line and
$\lambda^{''}_{312}=\lambda^{''}_{313}=\lambda^{''}_{323}=0.01$ 
for dashed line. 
In the calculations, we have neglected the decay width of 
$\tilde{t}_1$, since mass of $\tilde{t}_1$ is far from the
center-of-mass energy.
\begin{figure}[t]
\leavevmode
\begin{center}
\mbox{\epsfxsize=6.truecm\epsfysize=6.truecm\epsffile{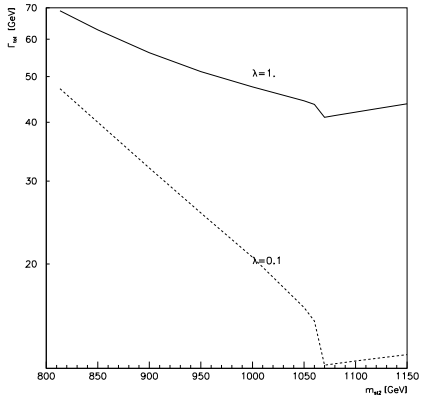}}
\end{center}
\caption{\label{stdecay}Decay width of stop quark $\tilde{t}_2$ 
               as function of mass of $\tilde{t}_2$. 
				  Solid line: $\lambda^{''}_{3ij}=1.0$.
        Dashed line: $\lambda^{''}_{3ij}=0.1$.}
\end{figure}
$\Gamma_{\tilde{t}_{2}}$ including $R_p$-violating contribution is shown
in Fig.5, where solid line corresponding to 
$\lambda^{''}_{312}=\lambda^{''}_{313}=\lambda^{''}_{323}=1$
and dashed line to 
$\lambda^{''}_{312}=\lambda^{''}_{313}=\lambda^{''}_{323}=0.1$.  
The results show that the cross sections 
can be very large if $m_{\tilde{g}}$ is smaller than $400$ GeV
and $\lambda^{''}$ are close to the present limits ($\lambda^{''} \sim 1$)
in the Tevatron Run II (with luminosity about $2\,{\rm fb}^{-1}$, 
it corresponds to hundreds of events). 
At LHC with luminosity $30\,{\rm fb}^{-1}$, the process can
potentially be seen with mass of gluino less than $1$ TeV even 
if $\lambda^{''} \le 0.01$.


We have also considered production of the LSP heavy gluino.
It has been shown in the model of \cite{s12} that masses of gluinos below
$115$ GeV are excluded except a narrow window
between $m_{\tilde{g}}=25 \ - \ 35$ GeV.
We found that the cross section of the process
in Tevatron Run I can be very large with masses of gluinos staying
in that narrow window (about $10$ pb with 
$m_{\tilde{g}}=30\, {\rm GeV}, \lambda^{''}_{3ij}=1$
and all $m_{\tilde{f}}=150$ GeV). 
Unlike in the discussion of \cite{s12},
gluinos can decay through $R_{p}$-violating interactions
(e.g. $\tilde{g}\rightarrow b c s$) if $\lambda^{''}$ 
is nonzero. 
The gluino decay added by single production of top-quark,
should have been visible already in Tevatron Run I using above results.  
Therefore, the narrow LSP gluino window can 
give much stronger constraints on $R_{p}$-violating
parameters $\lambda^{''}$ (with $\lambda^{''}=0.1$, still
several events are produced with luminosity $19\ pb^{-1}$ in Fermilab),
otherwise we can close this narrow LSP heavy gluino window. 

In the mSUGRA model, the gaugino soft-SUSY-breaking masses are
assumed to be equal at GUT scale and will lead to 
$m_{\tilde{g}}:m_{\tilde{\chi}^{\pm}_1}:m_{\tilde{\chi}^{0}_1}
\sim 7:2:1$ at low energy. However, there are other mechanisms
such as O-II model \cite{s15} in which ratios of masses can be given
as $m_3:m_2:m_1\sim -(3+\delta_{GS}):(1-\delta_{GS}):(33/5-\delta_{GS})$
at GUT scale, where $\delta_{GS}$ is the Green-Schwarz mixing term. 
Then at low energy,
we can obtain light
gluinos, which are almost degenerate with the lightest neutralino
and chargino, with the heavy scalar quarks. 
As a representative of
this model we take $\delta_{GS}=-4.1, \, m_0=1800\, 
{\rm  GeV}, m_{1/2}=10\, {\rm GeV}, A_{0}=0,
\tan \beta=3$ and $sign(\mu)=+1$ (leading to $m_{\tilde{g}}\sim 102$ GeV
and $m_{\tilde q}\sim 1.5$ TeV). 
Setting $\lambda^{''}_{3ij}=1$, the cross section of the process
$P\bar{P} \rightarrow (t+\bar{t}) \tilde{g} X$ can
be about $3.6$ fb with $\sqrt{s}=2$ TeV and of the process
$P{P} \rightarrow (t+\bar{t}) \tilde{g} X$
about $27$ pb with $\sqrt{s}=14$  TeV.
So in this model the single production of gluino
can provide a very significant signal for detecting SUSY and 
$R_{p}$ violation. Detecting pair production of gluino in this 
model was already discussed in \cite{s7,a1,s14} 
and it was shown that 
the $R_{p}$-violating decay of gluino
will dominate if $\lambda^{''}$ are close to the present
upper limits. Thus, the production of heavy quarks from gluino decay may 
provide a good signal for detecting gluinos.
Similarly, when we detect the production of $(t+\bar{t})\tilde{g}$,
we have at least two heavy quarks for tagging.  
 
\begin{flushleft} {\bf IV. Conclusion} \end{flushleft}
\par
We have studied the processes $PP(P\bar{P}) \rightarrow
(t+\bar{t}) \tilde{g} X$
in supersymmetric models with explicit $R_{p}$-violation.
We have seen that it is possible to test
the models at future Fermilab Tevatron Run-II and CERN LHC experiments, 
provided the $\lambda^{''}$-type couplings
are large enough.
We suggest also to check the old data in Tevatron Run I and get
stronger
constraints on $R_{p}$ violation or exclude the narrow window 
of LSP heavy gluinos.

The process, with single production of top quark
and gluino, can give signals both for SUSY and 
$R_{p}$ violation. 
Specifically, in a typical O-II
model, the process will be an important one
to check SUSY and $R_{p}$ violation. It is also shown
that gluinos should be detected through their $R_{p}$-violating
decay if $R_{p}$ violating parameters $\lambda^{''}$ are close
to the upper bounds.

\begin{flushleft} {\bf Acknowledgement} \end{flushleft}

Z.-H. Yu thanks the World Laboratory, Lausanne, for the scholarship.
  
\newpage
\begin{center} {\bf Appendix} \end{center}
\par
In this appendix we present the mass matrices \cite{s10} which we need in the
calculations.
\par
{\bf A. The sfermion sector}

In the mSUGRA model, the masses of squarks and sleptons follow from
the GUT scale parameters $m_{0}, m_{1/2}, A_{0}, \tan \beta$
and $sign(\mu)$ \cite{s10}. 
The sfermion mass matrices are given by
$$
\begin{array}{lll}
M_{\tilde{f}}^{2}= \left ( 
\begin{array}{cc} 
\ \ m_{\tilde{f}_{L}}^{2}+m_{f}^{2} \ \ \ \ \ \ \ m_{f} (A_{f}-\mu r_{f}) \\
\hspace{-0.5cm} m_{f} (A_{f}-\mu r_{f}) \ \ \ \ \ m_{\tilde{f}_{R}}^{2}+m_{f}^{2} 
\end{array} \right ),
\end{array}
\eqno{(A.1)}
$$
where $m_{\tilde{f}_{L,R}}$ are given in \cite{s10}, $m_{f}$ are
the masses of the partner fermions and 
$r_{d,s,b}=r_{e,\mu,\tau}=1/r_{u,c,t}=\tan \beta$. From these
matrices we can get the mixing angles $\theta_{f}$ and masses of the sfermions:
$$
\begin{array}{lll}
\sin 2\theta_{f} = \frac{2m_{f}(A_{f}-\mu r_{f})}{m_{\tilde{f}_{1}}^{2}-
m_{\tilde{f}_{2}}^{2}}\ ,\hspace{0.2cm} 
\cos 2\theta_{f} = \frac{m_{f_{L}}^{2}-m_{f_{R}}^{2}}{m_{\tilde{f}_{1}}^{2}-
m_{\tilde{f}_{2}}^{2}}, 
\end{array}
\eqno{(A.2)}
$$

$$
\begin{array}{lll}
m_{\tilde{f}_{1,2}}^{2}=m_{f}^{2}+\frac{1}{2} \left [ 
m_{f_{L}}^{2}+m_{f_{R}}^{2} \mp \sqrt{({m_{\tilde{f}_{L}}^{2}-
m_{\tilde{f}_{R}}^{2}})^{2}+4m_{f}^{2} (A_{f}-\mu r_{f})^{2}} \right ]. 
\end{array}
\eqno{(A.3)}
$$
\par
{\bf B. The chargino/neutralino sector}

In order to calculate the decay of gluino, we need to consider
the chargino and neutralino mass terms. 
The general chargino mass matrix is given as follows \cite{s10}:
$$
\begin{array}{lll}
M_{C}= \left ( 
\begin{array}{cc} 
\ \ m_{2} \ \ \ \ \ \ \ \sqrt{2} m_{w} \sin \beta \\
\hspace{-1.5cm} \sqrt{2} m_{w} \cos \beta \ \ \ \ \ \mu 
\end{array} \right ).
\end{array}
\eqno{(B.1)}
$$
It can be diagonalized by two real matrices U and V,
$$
\begin{array}{lll}
U^{*} M_{C}V^{-1},
\end{array}
\eqno{(B.2)}
$$
where $U=\cal {O}_{-}$, $V=\cal{O}_{+}$ if $detM_{C} \ge 0$ and
$V=\sigma_3 \cal{O}_{+}$ if $detM_{C} < 0$, with

$$
\begin{array}{lll}
\sigma_3= \left (
\begin{array}{cc} +1 \ \ 0\\0 \ \ -1 \end{array}
\right),
\cal{O}_{\pm}=
\left (
\begin{array}{cc} \cos \theta_{\pm} \ \ \sin \theta_{\pm}\\ -\sin \theta_{\pm} \ \ \cos \theta_{\pm} \end{array}
\right),
\end{array}
\eqno{(B.3)}
$$
and
$$
\begin{array}{lll}
\tan 2\theta_{-}=\frac{2\sqrt{2}m_{w}(m_2 \cos \beta+\mu \sin \beta)}{m_2^2-\mu^2-2m_w^2\cos \beta},\\
\tan 2\theta_{+}=\frac{2\sqrt{2}m_{w}(m_2 \sin \beta+\mu \cos \beta)}{m_2^2-\mu^2+2m_w^2\cos \beta}.\\
\end{array}
\eqno{(B.4)}
$$
Then the chargino masses are,
$$
\begin{array}{lll}
m_{x_{1,2}^{+}}&=& \frac{1}{\sqrt{2}}  [ m_2^2+\mu^2+2m_w^2\\
			&&\mp  \{ (m_2^2-\mu^2)^2+4m_w^4\cos^2 2\beta+4m_w^2(m_2^2+\mu^2+
2m_2\mu\sin 2\beta) \} ^{\frac{1}{2}}  ] ^{\frac{1}{2}}
\end{array}
\eqno{(B.5)}
$$
In the case of the neutralinos, the mass matrix is given by \cite{s10}:
$$
\begin{array}{lll}
M_{N}=\left (
\begin{array}{cc}
\hspace{0.5cm} m_{1}\ \ \ \ \hspace{2.0cm} 0 \ \hspace{1.2cm}  -m_z s_w \cos \beta \ \ m_z s_w \sin \beta\\
\hspace{0.5cm} 0 \ \ \ \hspace{2.0cm} \ m_2 \ \hspace{1.2cm}  m_z s_w \cos \beta \ \ -m_z s_w \sin \beta\\
\hspace{-0.5cm} -m_z s_w \sin \beta \ \ m_z c_w \sin \beta \ \hspace{1.2cm} 0 \hspace{2.0cm} \ \ \ \ -\mu\\
\hspace{-0.5cm} m_z s_w \sin \beta \ \ -m_z c_w \sin \beta \hspace{0.8cm}  -\mu \ \ \ \hspace{2.0cm} \ 0
\end{array}
\right ).
\end{array}
\eqno{(B.6)}
$$
It can be diagonalized by a unitary matrix $N$ ($4 \times 4$) as:
$$
\begin{array}{lll}
N^{T} M_{N} N.
\end{array}
\eqno{(B.7)}
$$

\end{document}